\documentclass[final,5p,times,10pt,sort&compress]{elsarticle}

\pdfoutput = 1
\usepackage{bm}
\usepackage{color}
\usepackage{graphicx}
\usepackage{amssymb}
\usepackage{cancel}
\usepackage{natbib}

{
{ 
{
\newcommand{\be}{\begin{equation}}
\newcommand{\ee}{\end{equation}}
\newcommand{\bea}{\begin{eqnarray}}
\newcommand{\eea}{\end{eqnarray}}

\newcommand{\bean}{\begin{eqnarray*}}
\newcommand{\eean}{\end{eqnarray*}}

\DeclareMathAlphabet{\mathpzc}{OT1}{pzc}{m}{it}

\journal{Physics Letter B}

\begin{document}

\begin{frontmatter}

\title{Co-genesis of Matter and Dark Matter with Vector-like Fourth Generation Leptons}

\author[UvA]{Chiara Arina}
\author[MCFP]{Rabindra N. Mohapatra}
\author[IIT]{Narendra Sahu}
\address[UvA]{GRAPPA Institute, University of Amsterdam, Science Park 904, 1090 GL Amsterdam, Netherlands}
\address[MCFP]{Maryland Center for Fundamental Physics and Department of Physics, University of Maryland, College Park, Maryland 20742, USA}
\address[IIT]{Department of Physics, IIT Hyderabad, Ordnance Factory Campus, Yeddumailaram 502 205, Medak, AP}

\begin{abstract}
We discuss aspects of a scenario for co-genesis of matter and dark matter  which extends the standard model by adding 
a fourth generation vector-like lepton doublet and show that 
if the fourth neutrino is a massive pseudo-Dirac fermion with mass in the few hundred GeV range and mass splitting of about 100 keV, its lighter component can be a viable inelastic dark matter candidate. Its relic abundance is produced by the CP violating out-of-equilibrium decay of the type-II seesaw scalar triplet, which also gives rise to the required baryon asymmetry of the Universe via type-II leptogenesis, thus providing a simultaneous explanation of dark matter and baryon abundance observed today. Moreover, the induced vacuum expectation value of the same scalar triplet is responsible for the sub-eV Majorana masses to the three active neutrinos. A stable fourth generation of neutrinos is elusive at collider, however might be detected by current dark matter direct search experiments.
\end{abstract}

\end{frontmatter}

\section{Introduction}\label{sec:intro} 
Dark Matter (DM), which constitutes 23\% of the total energy budget of the Universe is 
currently supported by the rotation curve of galaxies and clusters, gravitational lensing and large 
scale structure of the Universe. These indirect evidences suggest that the DM should be massive, 
electrically neutral and stable on the cosmological time scale~\cite{Bertone:2010zz}. The only information about DM 
hitherto known is its relic abundance which is precisely measured by the Wilkinson Microwave 
Anisotropy Probe (WMAP)~\cite{Komatsu:2010fb} and is given by $\Omega_{\rm DM} h^2=0.11$. However, the underlying mechanism which 
gives rise to the relic abundance is unknown. 

It is usually presumed that a weakly interacting massive particle of mass ${\cal O}(100)$ GeV can be 
a good candidate for DM as its annihilation cross-section $< \sigma|v| >\approx 3\times 10^{-26} 
{\rm cm}^3/{\rm s}$ satisfies the requirement of relic abundance, because it is produced by the standard thermal freeze-out 
mechanism~\cite{Kolb:1990vq}. However, an alternative mechanism has been explored in the literature, 
where the relic abundance of DM originates  via the asymmetric component rather than the symmetric 
component of any stable species. In this case, the relic abundance depends on the amount of 
CP-violation in the theory, in a similar way to the baryogenesis mechanism~\cite{Dodelson:1991iv,Kaplan:1991ah,Kuzmin:1996he,Fujii:2002aj,Oaknin:2003uv,Hooper:2004dc,Kitano:2004sv,Cosme:2005sb,
Farrar:2005zd,Roszkowski:2006kw,McDonald:2006if,Kaplan:2009ag,Kohri:2009yn,An:2009vq,Frandsen:2010yj,
Feldstein:2010xe,An:2010kc,Cohen:2010kn,Shelton:2010ta,Davoudiasl:2010am,Haba:2010bm,Gu:2010ft,Blennow:2010qp,
McDonald:2011zz,Dutta:2010va,Haba:2011uz,Falkowski:2011xh,Chun:2011cc,
Buckley:2011kk,Graesser:2011wi,Iminniyaz:2011yp,Heckman:2011sw,MarchRussell:2011fi,Davoudiasl:2012uw,MarchRussell:2012hi,Gu:2012fg}.    
 
In this article we study the possibility of adding a vector-like lepton doublet to the standard model (SM) whose neutral member (to be called fourth neutrino henceforth) could be a candidate for  DM. Indeed a fourth generation of fermions~\cite{Mohapatra:1993vm,Holdom:2009rf,Joglekar:2012vc,Lee:2012xn,Frampton:1999xi,Geller:2012tg,An:2012vp} is one of the simplest 
extension of physics beyond the SM with rich phenomenology and also extensively searched for at colliders. The properties 
of the new family are subject to tight constraints from electroweak precision measurements and by direct 
searches \cite{Erler:2010sk,Beringer:1900zz}. Considering the fourth generation leptons, probably the most stringent bound is the $Z$ invisible width measured at LEP, because it provides strong evidence for only three family of light neutrinos. A 4th generation neutrino, if present, should be very distinct in nature from the three SM neutrinos. Indeed it should be heavier than at least $m_Z/2$, in order to avoid conflict with $Z$ decay width measurement. Therefore the model of fourth generation leptons we present is distinct from the idea of sequential repetition of the SM fermionic families.

As is well-known, in simple  heavy fourth generation extensions of SM, the heavy neutrino ($N_4$), which is part of a lepton doublet $L_4\equiv (N_4, E_4)$, does not qualify as a dark matter since rapid $N_4\bar{N}_4$ annihilation to SM particles via $Z$-exchange reduces  its relic density  to a value far below what is required for it to be a viable DM candidate as well as is excluded by direct DM searches due to its coupling with the $Z$ boson. Our model for the fourth generation neutrino $N_4$ is however different:  in addition to being part of a vector-like doublet, it has two additional features, which endow it with the properties that  make it a viable dark matter candidate. (i) $N_4$  is a pseudo-Dirac neutrino, whose Majorana mass arises from the vev (vacuum expectation value) of a $Y=2$ Higgs triplet $\Delta$, acquired below electroweak ($wk$) phase transition.  We will call this the type-II seesaw Higgs field, which anyway is present in our model to make the familiar active neutrinos acquire mass via the type-II seesaw mechanism. The presence of this Majorana mass makes it an inelastic
 dark matter~\cite{TuckerSmith:2001hy}, that has the advantage of fitting the results of current DM search experiments and not being excluded by upper limits. To keep the fourth family lepton doublet stable, we then impose an extra $Z_2$ symmetry on the model under which the fourth family lepton doublet $L_4$ is odd and all other fields of the theory are even~\cite{Arina:2011cu,Arina:2012fb}: besides the fourth family neutrino being lighter than the corresponding charged lepton, it is decoupled from the other lepton doublets. (ii) Secondly, the decay of the two type-II seesaw Higgs triplets via their CP violating coupling produces an asymmetry in the fourth family lepton number, which is large enough so that the depletion problem of relic density alluded to above does not occur. In fact, this asymmetry is comparable to the ordinary lepton number generated in the same decay which gives rise to the matter anti-matter asymmetry in the Universe via leptogenesis~\cite{Arina:2011cu,Arina:2012fb}. Both asymmetries can be comparable to each other in realistic models. In other words, the triplet mass scale is superheavy so that its CP violating out-of-equilibrium decay can produce asymmetry simultaneously in the DM and lepton sector and above the electroweak phase transition temperature, the lepton asymmetry for  the familiar leptons gets converted to the baryon asymmetry via $SU(2)_L$ sphalerons~\cite{Harvey:1990qw}. In this case, we want to emphasize that the generated lepton asymmetry in the fourth generation does not get converted to baryon via sphaleron processes since $L_4$ being a vector-like doublet, it does not contribute to the B+L anomaly of the standard model. On the other hand the symmetric component gets depleted via rapid annihilation, {\it i.e.} $Z$-exchange. The common origin of two asymmetries from the $\Delta$ decay then naturally explain the similar order of magnitude for the DM-to-baryon ratio and by adjusting the masses and couplings in both sectors, one can have $\Omega_{\rm DM}/\Omega_B \sim 5$. Thus our model provides another example of co-genesis of matter and dark matter. 
 
It is worth mentioning that in this paper we focus on the model building aspects of the co-genesis mechanism with respect to Ref.~\cite{Arina:2011cu,Arina:2012fb} and try to address some issues about the viability of the scenario described above that were left unexplored. In particular we propose a mechanism to introduce a splitting in mass between the neutral and charged partner of the vector-like doublet and we investigate the survival of the asymmetry at electroweak phase transition. Lastly we update the direct detection part with the latest data release by XENON100~\cite{Aprile:2012nq}, investigate if the model might accommodate the excess seen by the CRESST-II detector~\cite{Angloher:2011uu} and if there is a compatibility with the KIMS exclusion bound~\cite{Kim:2012rz}. 

Our letter is organised as follows. In section~\ref{sec:pot} we present the model for a fourth generation of fermions, 
discussing in section~\ref{sec:ewconst} constraints from electroweak precision measurements and direct searches at colliders. 
The phenomenology for generating the asymmetries and the measured DM-to-baryon ratio is presented in section~\ref{sec:pheno} together with the constraints from DM direct searches. We then summarize in section~\ref{sec:concl}.

\section{Fourth Generation Pseudo-Dirac Neutrino as DM}\label{sec:pot}

Fourth family neutrino has been studied as a dark matter in gauge extensions of the standard model by several authors~\cite{Lee:2011jk,Zhou:2011fr,
Joglekar:2012vc,Lee:2012xn}. In this study, we focus on a vector like 4th generation lepton doublet, $L_4$, which will 
give a candidate of inelastic DM and being vector-like will not need the new set of quarks for anomaly cancellation.

\subsection{Triplet Seesaw and Sub-eV Majorana Masses of Three Active Neutrinos}

In addition to the vector-like lepton doublet, we add two scalar triplets $\Delta_{1,2}$ with $Y=2$. Since the hypercharge of $\Delta$ is 2, it can have bilinear coupling to Higgs doublet $H$ as well as to the lepton doublets. The scalar potential involving $\Delta$ (from here on we drop the subscripts for the two scalar triplets and refer to them loosely as $\Delta$) and $H$ can be written as follows:
\begin{eqnarray}\label{eq:ScalarPotential}
\hspace*{-0.5cm}
 V(\Delta, H) & = &   M_\Delta^2 \Delta^\dagger \Delta + \frac{\lambda_\Delta}{2} (\Delta^\dagger \Delta)^2 -  M_H^2 H^\dagger H \nonumber\\
&+&  \frac{\lambda_H}{2} (H^\dagger H)^2 + \lambda_{\Delta H} H^{\dagger} H \Delta^\dagger \Delta \nonumber\\
&+& \frac{1}{\sqrt{2}} \left[ \mu_H \Delta^\dagger H H + {\rm h.c.}\right]  \,.
\end{eqnarray}
The bi-linear couplings of leptons and Higgs to scalar triplet are given by:
\begin{equation}\label{eq:Lag-DM}
\hspace*{-0.5cm}
 -\mathcal{L}  \supset  \frac{1}{\sqrt{2}} \left[ f_H M_\Delta \Delta^\dagger H H + (f_L)_{\alpha,\beta} \Delta 
L_\alpha L_\beta + {\rm h.c.} \right]\,,
\end{equation}
where $f_H=\mu_H/M_\Delta$ and $\alpha, \beta=1,2,3$. Below electroweak phase transition the scalar triplet acquires an induced vev:
\begin{equation}
\langle \Delta \rangle = -f_H \frac{v^2}{\sqrt{2} M_\Delta}\,,
\end{equation}
where $v=\langle H \rangle$ = 246 GeV. The value of $\langle \Delta \rangle$ is upper bounded to be around 1 GeV in 
order not to spoil the SM prediction: $\rho \approx 1$. The $\Delta L_\alpha  L_\beta$ coupling gives Majorana masses to three flavors of active neutrinos as:
\begin{equation}
(M_\nu)_{\alpha \beta}= \sqrt{2} f_{\alpha\beta}\langle \Delta \rangle = -f_{L,\alpha\beta}f_H \frac{v^2}{\sqrt{2} M_\Delta}\,.
\end{equation}
Taking $M_\Delta \sim 10^{10}$ GeV, $f_H\sim 1$ and $f_L \sim {\mathcal O}(10^{-4})$ we get $M_\nu \sim {\mathcal O}$(eV), which is 
compatible with the observed neutrino oscillation data \cite{Fukuda:2001nk,Ahmad:2002ka,Eguchi:2003gg}.

%%%%%%%%%%%%%%%%%%%%%%%%%%%%%%%%%%%%%%%%%%%%%%%%%%%%%%%%%%%%%%%%%%%%%%%%%%%%%%%%%%%%%%%%%%%%%

\subsection{Triplet Seesaw and Pseudo-Dirac mass of fourth Generation Neutrino}
%%%%%%%%%%%%%%%%%%%%%%%%%%%%%%%%%%%%%%%%%%%%%%%%%%%%%%%%%%%%%%%%%%%%%%%%%%%%
The Lagrangian that gives the 4th family neutrino its mass is given by:   
\begin{equation}
-\mathcal{L}_{L_4-mass} = M_D\overline{L_4} L_4 + \frac{f_4}{\sqrt{2}}\overline{L_4^c}i\tau_2 \Delta L_4 + {\rm h.c.}
\label{eq:nu4mass}
\end{equation}
where $M_D$ generates the Dirac mass of the $N_4$. Below electroweak phase transition $\Delta$ acquires an induced vev and generates a Majorana mass $m=\sqrt{2}f_4\langle \Delta \rangle$ for $N_4$. Therefore, the Dirac spinor $N_4$ can be written as sum of two Majorana spinors $(N_{4,L})$ and $(N_{4,R})$. As a result the Lagrangian (\ref{eq:nu4mass}) becomes:
\begin{eqnarray}
-\mathcal{L}_{L_4-mass} & = &  M_D \left[\overline{(N_{4,L}) } ((N_{4,R})+ \overline{((N_{4,R}) } ((N_{4,L}) \right]\nonumber\\
& + & m \left[ \overline{(N_{4,L})^c} (N_{4,L}) + \overline{(N_{4,R})^c} (N_{4,R})  \right]\,.
\end{eqnarray}
This implies that there is a $2\times 2$ mass matrix for the fourth generation neutrino in the basis 
$\{N_{4,L}, N_{4,R} \}$. By diagonalising the mass matrix we get the two mass eigenstates $N_1$ and $N_2$ with mass eigenvalues $(M_D-m)$ and $(M_D+m)$. Thus the mass splitting between the two 
states is given by:  
\begin{equation}\label{eq:delta}
\delta= 2m = 2 \sqrt{2}f_4\langle \Delta \rangle\,. 
\end{equation}
We assume that the mass splitting is small, namely $\delta \sim {\cal O}(100)$ keV, compared to the mass scale of these states, which is of 
order 100 GeV. Therefore, the two mass eigenstates are pseudo-Dirac type neutrino and act as inelastic DM. The lighter of them is indeed stable, because of the discrete $Z_2$ symmetry we imposed. Besides the fourth generation being inert, namely it does not couple to the three SM families of fermions, it does not couple neither to the 
Higgs boson, implying that all the Yukawa couplings to the SM Higgs field are zero. The masses of the vector-like 4th generation are therefore not linked to 
electroweak symmetry breaking and are not predicted by the model. We however suppose them at the electroweak scale and take into account the constraints from LEP direct searches.

\subsection{Mass splitting between the charged and neutral component of $L_4$ } 
An important part of the discussion of dark matter neutrino in our model is the splitting between the charged and the neutral member of the fourth generation lepton doublet. A simple way to achieve this without disturbing other aspects of the model is to introduce an SM singlet lepton $N$ with near TeV scale mass $M_N$ and additional Higgs doublet $H^\prime$, with yukawa couplings of the order of $\mathcal{O}(0.1-1)$. The extra fields transform under the $Z_2$ as $L_4\to -L_4$, $H^\prime \to H^\prime$ and $N\to - N$. 
Once $H^\prime$ acquires a vev $v^\prime_{wk}$, the $N_4$ and $N$ field get a $2\times 2$ mass matrix of the form:
\begin{eqnarray}
M_{N_4,N}~=~\left(\begin{array}{cc}M_4 & h^\prime v^\prime_{wk}\\ h^\prime v^\prime_{wk} & M_N\end{array}\right)
\end{eqnarray}
This lowers the mass of the dark matter neutrino to the value $m_{N_4}\equiv M_{\rm  DM}\sim M_4-\Delta m \sim M_{4}-\frac{(h^\prime v^\prime_{wk})^2}{M_N}$.

%%%%%%%%%%%%%%%%%%%%%%%%%%%%%%%%%%%%%%%%%%%%%%%%%%%%%%%%%%%%%%%%%%%%%%%%%%%%%%%%%%%%%%

\section{Pseudo-Dirac fourth generation neutrino as dark matter}\label{sec:pheno}

\subsection{Co-genesis of matter and dark matter}

Since the scalar triplet is superheavy, it decays in the early Universe in a quasi-equilibrium state in various channels, 
namely $\Delta\to L_\alpha L_\beta$, $\Delta \to L_4 L_4$ and $\Delta \to HH$. The decay channels can 
be easily read from the Lagrangian (\ref{eq:Lag-DM}). Since these couplings are in general complex, charge conjugation (C) and 
parity (P) are jointly violated through the interference of tree-level and one loop self-energy correction diagrams. As a result 
the decay of $\Delta$ produces asymmetries in the visible ({\it i.e.} $\Delta \to L_\alpha L_\beta$) sector and in the DM sector 
({\it i.e.} $\Delta \to L_4 L_4$). The asymmetry in the Higgs disappears after the later acquires a vev. However, the asymmetries in the 
visible and DM sectors remain forever. 

\begin{figure}
\centering
\includegraphics[width=1.03\columnwidth,trim=22mm 70mm 30mm 75mm, clip]{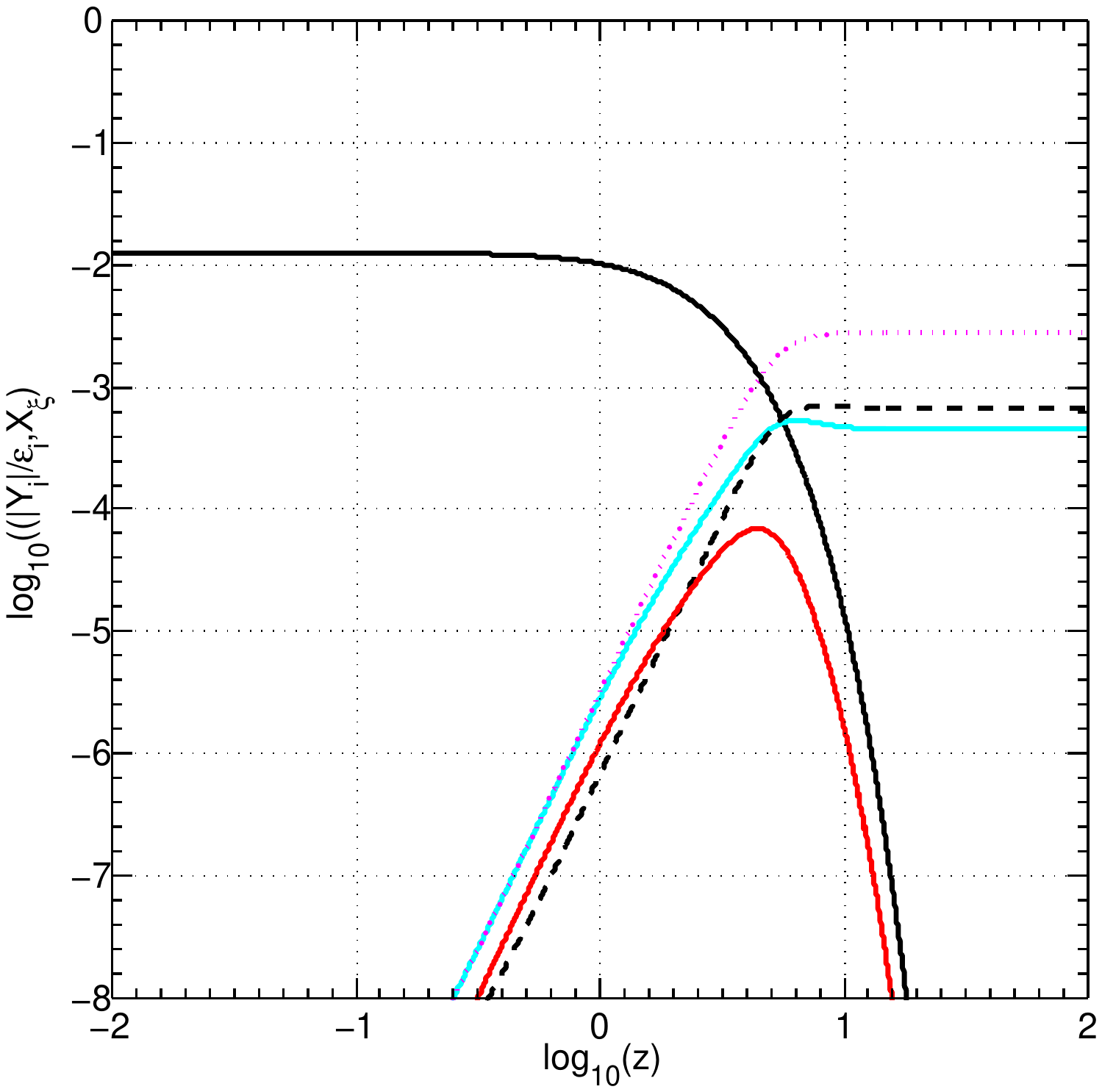}
\caption{Absolute value for the Yield of leptons (cyan solid), DM (dotted magenta), Higgs (dashed black), scalar triplet asymmetry (solid red) plus scalar triplet abundance (black solid), for a successful point with $m_{\rm DM} = 60$ GeV, $B_L =0.015$, $B_{\rm DM} =1.7\times10^{-5}$, $\epsilon_L =3.4\times 10^{-7}$, $\epsilon_{\rm DM} =3.6 \times 10^{-8}$, which leads to $\Omega_{\rm DM}/\Omega_B = 5.0$, $Y_L = 1.6\times 10^{-10}$, $Y_{\rm DM} = 1.0\times 10^{-10}$ and $\eta_{\rm DM}/\eta_L = 0.48$.}\label{fig:ex}
\end{figure}

Quantitatively, the asymmetries in the lepton and dark matter sectors are as follows
\begin{eqnarray}
& Y_L \equiv \frac{n_L}{s} =  \epsilon_L X_\Delta \eta_L\,,\\ 
& Y_{\rm DM} \equiv \frac{n_{L_4}}{s} = \epsilon_{L_4} X_\Delta \eta_{L_4}\,,
\label{DM-asy}
\end{eqnarray} 
where $X_\Delta=n_\Delta/s$, with $s=(2\pi^2/45) g_* T^3$ the entropy density and $n_\Delta$ the number density of the triplet scalar. $\eta_L$, $\eta_{L_4}$ are the 
efficiency factors which take into account the depletion of asymmetries due to the number violating processes involving 
$L_\alpha, L_4$ and $H$. At a temperature above electroweak phase transition the lepton asymmetry gets converted to 
baryon asymmetry via the $SU(2)_L$ sphalerons as:
\begin{equation}
Y_B= -0.55 Y_L\,.
\label{B-asy}
\end{equation}
As noted in~\cite{Arina:2011cu,Arina:2012fb}, the primordial $L_4$ asymmetry is much larger than the primordial value of the familiar lepton asymmetry by a factor of $f^2_H/f^2_L$ (nearly $10^8$). The enhanced annihilation rate of $L_4$ causes a much stronger wash-out of $\epsilon_{L_4}$ via the processes $L_4L_4\to HH$ than of the corresponding asymmetry $\epsilon_L$ for familiar leptons, whose couplings are much smaller. Using this and the equations (\ref{DM-asy}) and (\ref{B-asy}), we get the DM to baryon abundance:
\begin{equation}
\label{eq:ratio}
\frac{\Omega_{DM}}{\Omega_B} = \frac{1}{0.55} \frac{m_{N_4}}{m_p} \frac{\epsilon_{L_4}}{\epsilon_L} \frac{\eta_{L_4}}{\eta_L}\,,
\end{equation}
where $m_p \sim 1$ GeV is the proton mass and $\eta_{L_4, L}$ represent the wash out effect. The details of the numerics can be found in 
references~\cite{Arina:2011cu,Arina:2012fb}, where a phenomenological analysis of the parameter space satisfying $\Omega_{\rm DM} \sim 5\Omega_B$ has been realized. Here we plot in figure~\ref{fig:ex} a particular solution for the co-genesis mechanism: we observe that the asymmetry generated in the DM sector ($Y_{\rm DM} =1.0 \times 10^{-10}$) is of the same order of the asymmetry in the leptons ($Y_L = 1.6 \times 10^{-10}$) and hence in the baryonic sector. The efficiency in the dark matter channel is although larger than the efficiency in the leptonic channel because it should compensate the effect of a large DM mass (see equation~\ref{eq:ratio}) and a small CP asymmetry; the fast channel is the Higgs one. The parameters used for the solution of the Boltzmann equations as well as the absolute yields are given in the caption and are representative of a large portion of the allowed parameter space. Viable solutions can be find for dark matter masses running up to TeV scale, even though they are disfavoured with respect to solutions at lower dark matter mass because of the naturalness principle: since the ratio of DM to baryon abundance is close to unit it is more natural to have light dark matter with the same efficiency and CP asymmetries than the visible matter. Larger dark matter masses are allowed because of the compensation effect between asymmetries and efficiency factors, as described by equation~\ref{eq:ratio}.

We wish to point out that it is possible to construct theories where the two Higgs triplets couple to the different set of leptons (one to familiar ones and the other to $L_4$) due to the existence of some symmetry but mix with each other with a small mixing after symmetry breaking. In this case, the hierarchy between $f_H$ and $f_L$ can be of order $10^{-2}$ or so, so that the ratio between $\frac{\epsilon_{L_4}}{\epsilon_L}$ is much less than in the model described above. There can be a larger range of parameters where current dark matter abundance can be fitted. However, in this case the concept of co-genesis has to be sacrificed at the leading order.

\subsection{Cosmological evolution of dark matter below electroweak phase transition} 

As emphasized in the previous section, even though the primordial $L_4$ (DM) asymmetry is much larger than the familiar lepton asymmetry, strong wash-out effective above the electroweak phase transition epoch $T=T_{wk}$, brings them to be of similar magnitude. An important issue arises after electroweak phase transition, when there is the small Majorana mass for $L_4$ which turns on below $T_{wk}$. This splits the $L_4$ into two Majorana eigenstates $N_1$ and $N_2$ by 100 keV mass. The question to be addressed now is: can the two states annihilate to reduce $\Omega_{DM}$? As it has been noted in~\cite{Arina:2011cu}, if the DM mass is $\geq 2$ TeV, $L_4\bar{L}_4$ annihilation freezes out before $T_{wk}$ and no further reduction of $\Omega_{DM}$ takes place. However, what happens for lower masses needs to be discussed, {\it i.e.} do we lose the $L_4$ asymmetry via weak annihilation processes below $T_{wk}$. 

There are two possible things that can happen: the two Majorana eigenstates can annihilate each other via both the lepton number conserving and the lepton number violating processes, where the latter involves the Majorana mass $\delta /2$. The dominant lepton number conserving annihilation only reduces the symmetric component  but not the asymmetric part which would require the intervention of the small Majorana mass $\delta/2$. Since relic density of DM is due to the asymmetric part,  if the $L_4$ violating reaction rates are out of equilibrium, in this range, the ``turning on'' of $\delta/2$ will not affect the relic density. We therefore give a heuristic discussion of whether this is the case. We expect the $L_4$ violating part of the annihilation to be proportional to the parameter $\delta/2$.

\begin{figure}
\centering
\includegraphics[width=1.03\columnwidth,trim=18mm 70mm 33mm 75mm, clip]{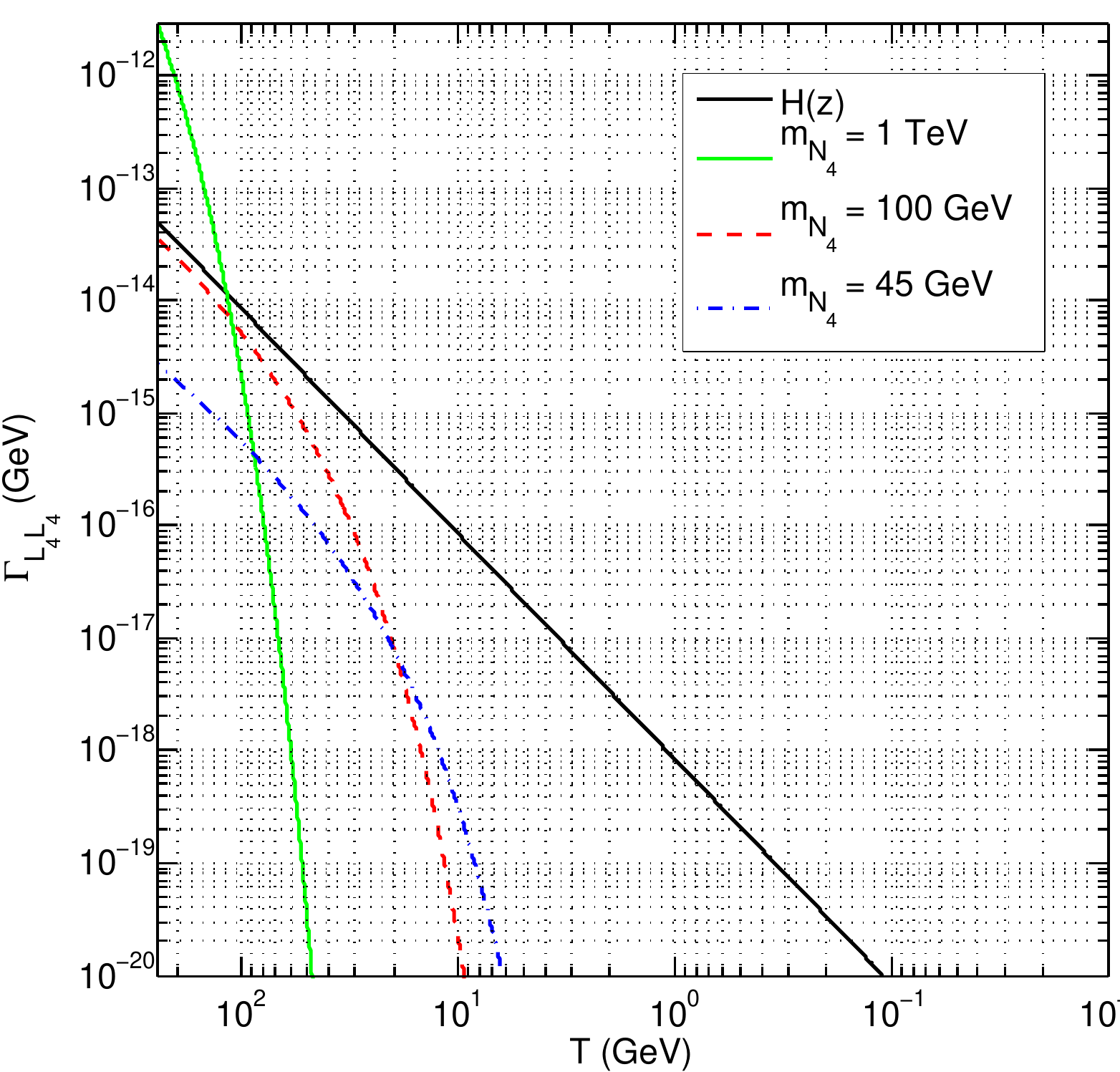}
\caption{The scattering rate of the process $L_4L_4\to f\bar{f}$ as a function of the temperature is compared with the Hubble expansion rate. For illustration purpose we have assumed the Majorana mass splitting to be 100 keV and we have considered three values for the mass of the fourth generation neutrino as labelled.}\label{fig:washout}
\end{figure}

In order to give a qualitative ``feel'' for the above argument, we note that the rate for the lepton number depleting process, $\Gamma ( L_4L_4\to f\bar{f})$ via $Z$-exchange is expected to be given by:
\begin{eqnarray}
\Gamma ( L_4L_4\to f\bar{f})\simeq \frac{G^2_F M^2_D}{2\pi} c_{\theta_W} \left(\frac{\delta}{2 T}\right)^2\frac{n_{L_4}}{n_\gamma}T^3\,,
\end{eqnarray}
where $G_F$ is the Fermi coupling constant, $c_{\theta_W}$ the cosine of the Weinberg angle and we have used the Boltzmann distribution to account for the non-relativistic number density of $N_4$ particles. As a result below $T_{wk}$, we find that this lepton number depletion rate suffers an exponential suppression and 
therefore it is slower than the expansion rate of the universe for the range of masses we are interested in. Hence this process is not very effective in reducing the dark matter asymmetry, as shown in figure~\ref{fig:washout}. We therefore believe that once the dark matter asymmetry has been created above $T_{wk}$, it will survive till the present epoch.

Another issue is the possible oscillation of $N_4\to \bar{N}_4$ via the $\delta/2$~\cite{Arina:2011cu,Buckley:2011ye,Cirelli:2011ac,Cui:2009xq} below the temperature when triplet vev turns on. Note that if the Majorana mass turns on below the freeze-out temperature for $N_4\bar{N}_4$ annihilation, the oscillations simply redistributes the relic density between $N_1$ and $N_2$ and when $N_2$ decays to $N_1$, the net relic density remains unchanged. This is for example the case when $M_{\rm DM}\geq 2$ TeV. If $M_{\rm DM}\leq 2$ TeV, there are two possibilities:

(i) Unlike generic DM, our DM candidate has weak  as well as magnetic moment interactions with the hot plasma of the early universe. Discussion of such oscillations in the presence of dense medium as the early Universe is not very simple~\cite{Akhmedov:2012mk} and it is not clear how to estimate the oscillation rate in such a situation. We therefore assume that such oscillations do not play an important role in depleting the $N_4$ asymmetry for $M_{\rm DM}\leq 2$ TeV. 

(ii) Second possibility is to modify the model such that the Majorana mass  arises due to a triplet vev ``turning on''  at a much lower temperature than $T_{wk}$. For example, we could consider multi-Higgs doublet models with the Higgs fields that couple to $\Delta$ to induce triplet vevs themselves have vevs of order of a few GeVs (as in high $\tan\beta$  two Higgs models). This would require $\mu_\Delta \gg M_\Delta$ ( e.g. $\mu_\Delta\sim 10^{13}$ GeV and $M_\Delta \sim 10^9$ GeV). In such models, the Majorana mass $\delta/2$ will turn on around 5 GeV so that we could allow $M_{\rm DM}\geq100$ GeV  and for such masses, by the time $\delta/2$ turns on, the $N_4$ freeze-out would have taken place and as we argued before, the relic density will not be reduced further.

\subsection{Fourth Generation Neutrino and DM Direct Searches}
\begin{figure}
\centering
\includegraphics[width=1.\columnwidth,trim=28mm 64mm 35mm 67mm, clip]{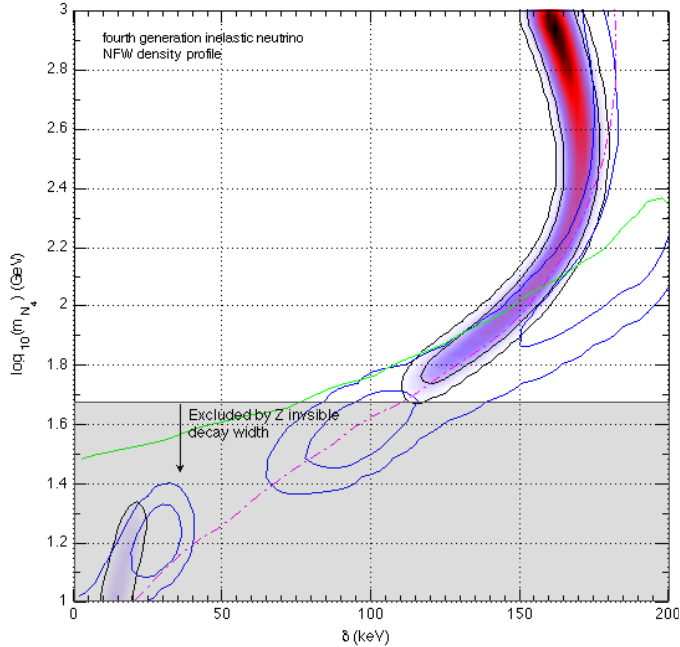}
\caption{2D marginal posterior pdf in the $\{\delta,m_{N_4}\}$-plane. The shaded (blue solid) contours denote the 90\% and 99\% credible regions for DAMA (CRESST) respectively. The magenta dot-dashed line is the XENON100 exclusion limit, while the green dashed line is the upper bound of KIMS experiment, at $90_S\%$ confidence level~\cite{Arina:2012dr}. All the astrophysical uncertainties and nuisance parameters have been marginalized over. The light gray region is excluded by LEP.}\label{fig:dd}
\end{figure}

We now make a few comments on the implications of our model for dark matter search. As noted, the coupling between $N_4$ and $\Delta$ provides a small Majorana mass to the 4th generation of neutrinos. In the mass basis, $N_1$ has an off diagonal coupling with the $Z$ boson,  preventing it to be excluded by direct detection searches. If the mass splitting is of the order of several keV, the DM $N_1$ actually has enough energy to scatter off nuclei and to go into its excited state $N_2$, which is the definition of inelastic scattering~\cite{TuckerSmith:2001hy}. 

The state of art for a 4th generation inelastic neutrino is given by figure~\ref{fig:dd} in the $\{\delta, m_{N_4}\}$-plane, where the cross-section is fixed by the model, while the Majorana mass is allowed to vary in a reasonable range of values, in order for the scattering to occur. A Majorana mass of the order of 100 keV accounts for the DAMA~\cite{Bernabei:2010mq} annual modulated signal (shaded region), while a much wider range accounts for the event excess seen in CRESST~\cite{Angloher:2011uu} (blue non filled region). However those regions are severely constrained by XENON100~\cite{Aprile:2012nq} and KIMS~\cite{Kim:2012rz}. KIMS is very constraining being a scintillator with Iodine crystals as DAMA. Our dark matter candidate can explain simultaneously the DAMA and CRESST detection, with a  marginal compatibility at $90_S\%$ with XENON100 and KIMS, for a mass range that goes from 45 GeV up to $\sim 250$ GeV. If we give up the DAMA explanation, then it could account for the CRESST excess up to masses of the order of $\sim 500$ GeV. 

The details on the model cross-section are given in~\cite{Arina:2011cu}, while for the numerical analysis of the latest experimental results we refer to~\cite{Arina:2012dr}.

\section{Electroweak Precision Tests and Direct Limits on Fourth Generation Leptons}\label{sec:ewconst}
%%%%%%%%%%%%%%%%%%%%%%%%%%%%%%%%%%%%%%%%%%%%%%%%%%%%%%%%%%%%%%%%%%%%%%%%%%%%%%%%%%%%%%%%%%%%%%%%%%%%%%%%%%%%%%
Nowadays a fourth family of fermions, in particular chiral and whose mass is related to electroweak symmetry breaking, is very 
severely constrained by LHC with the Higgs-like signal at 125 GeV, flavour violating processes and electroweak precision tests~\cite{Buchkremer:2012yy,Eberhardt:2012gv,BarShalom:2012ms,Bellantoni:2012ag,Djouadi:2012ae}, perhaps almost ruled out. One of the reasons 
is that the 4th generation of quarks modifies the production of the Higgs boson and depletes the $h \to \gamma \gamma $ decay channel, which 
goes into contradiction with the experimental data. However the constraints on a 4th generation of fermions strongly depend on the assumptions 
of the model~\cite{Erler:2010sk}. For example it has been shown that vector-like families can provide the measured branching ratio for $h \to \gamma\gamma$ 
and be compatible with electroweak precision measurements~\cite{Joglekar:2012vc}.

\begin{figure}
\centering
\includegraphics[width=0.9\columnwidth,trim=0mm 0mm 0mm 0mm, clip]{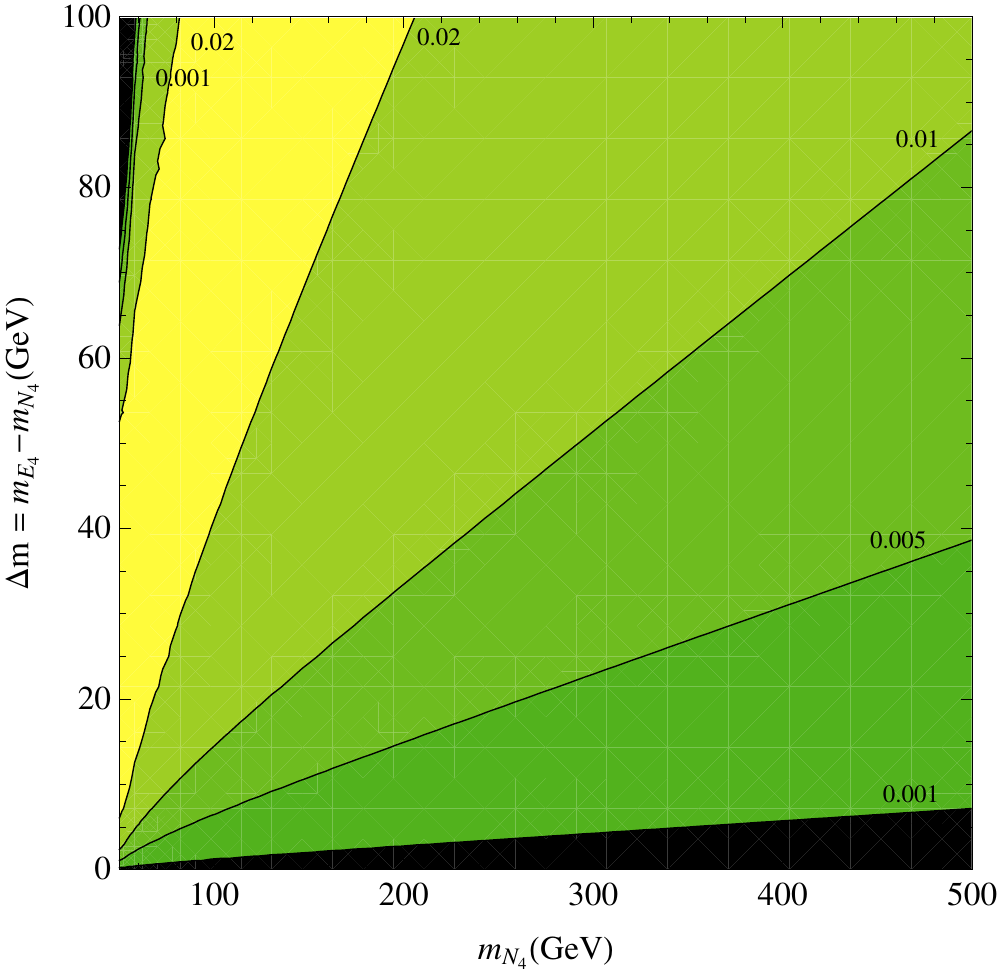}
\caption{Contour plot for the oblique corrections to $\Delta S$ in the plane $\{m_{N_4},\Delta m\}$. The black solid lines indicate some reference values for $\Delta S$ as a function of the 4th generation neutrino mass and of the lepton doublet mass splitting, as labelled.}\label{fig:ds}
\end{figure}

\begin{figure}
\centering
\includegraphics[width=0.9\columnwidth,trim=0mm 0mm 0mm 0mm, clip]{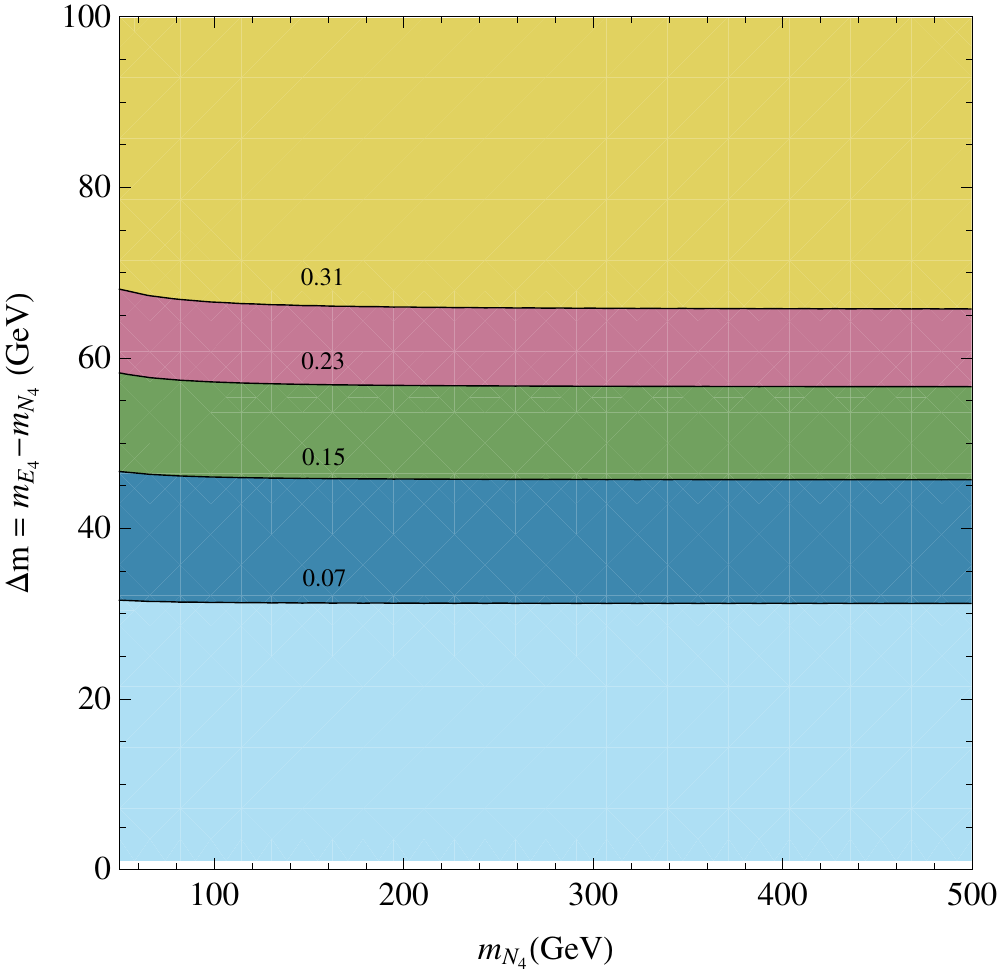}
\caption{Same as figure~\ref{fig:ds} for the oblique corrections to $\Delta T$. As labelled, the black solid lines indicate the central value as well as the $1,2,3\ \sigma$ contours.}\label{fig:dt}
\end{figure}

If really the Yukawa couplings between $L_4$ and $H$ are zero as well, as in our model, the only constraints come from the oblique parameters S and T~\cite{Peskin:1991sw} and 
from direct measurements at LEP. These latter are as follows: the $N_4$ are pseudo-Dirac neutrinos and are stable, hence lower bounded by the invisible $Z$-decay width, which gives $m_{N_4} > 45 \, {\rm GeV}$. The bound on the mass can not be lowered  for the Majorana case~\cite{Carpenter:2010dt}  because it relies on the process $Z \to N_4 \bar{N}_4$ which contributes only to the invisible width of the $Z$ boson. However, the charged partner $E_4^{\pm}$ can be searched for in the collider and is required  to be heavier than $N_4$. In particular, the pair production of $E_4^-E_4^+$ at LEP with subsequent decay to SM particles and missing energy (in the form of neutrino and DM) puts a 
lower limit on its mass scale to be~\cite{Beringer:1900zz}:
\begin{equation}
m_{E_4}>101.9\,  {\rm GeV}~~{\rm and}~~m_{E_4} -  m_{N_4} \equiv \Delta m > 15\,  {\rm GeV}.
\end{equation}

The effects of new physics, which does not necessarily couples to SM fermions, manifest in the $W$ and $Z$ boson self-energies and are measured by the 
corrections to oblique parameters $S$, $T$ and $U$. Those parameters are well constrained by electroweak precision data and the allowed deviations from the SM 
model are~\cite{Beringer:1900zz}:
\begin{equation}
\Delta S = 0.04 \pm 0.09 \, \, \, {\rm and} \, \, \, \Delta T = 0.07 \pm 0.08
\end{equation}
with $\Delta U = 0$, which is a good assumption because the oblique contribution from a 4th generation to $\Delta U$ is negligible.

For a fourth generation of vector-like leptonic doublet the oblique corrections are given by:
\begin{eqnarray}
\Delta S & = &\frac{1}{\pi}\left[\frac{22 y_1 + 14 y_2}{9} \frac{1}{9}\ln\frac{y_1}{y_2}+\frac{11 y_1 + 1}{18} f(y_1)\right.\nonumber\\
& + & \left.\frac{7 y_2-1}{8}f(y_2)-\sqrt{y_1 y_2}\left(4+ \frac{f(y_1)f(y_2)}{2}\right)\right] \,,\nonumber\\
\Delta T & = & \frac{1}{8 \pi s^2_{\theta_W} c^2_{\theta_W}} \left[y_1 + y_2 - \frac{2 y_1 y_2}{y_1-y_2}\ln \frac{y_1}{y_2} \right.\nonumber\\
& + & \left. 2 \sqrt{y_1 y_2}\left( \frac{y_1+y_2}{y_1-y_2} \ln \frac{y_1}{y_2}-2\right)\right]\,,
\end{eqnarray}
having defined $y_i = m^2_i/m^2_Z$ while $s^2_{\theta_W}$ is the sine square of the Weinberg angle. The mass 
term $m_i$ refers to the mass of the 4th generation of leptons. The function $f(y_i)$ is defined as:
\begin{eqnarray}
\hspace*{-0.8cm}
 f(y_i) \equiv \left\{
\begin{array}{ll}
-2 \sqrt{\Delta(y_i)} \left(\arctan \frac{1}{\sqrt{\Delta(y_i)}}-\arctan \frac{-1}{\sqrt{\Delta(y_i)}}\right)& \Delta(y_i)  > 0\,,\nonumber\\
0 & \Delta(y_i) =0\,,\nonumber \\
\sqrt{-\Delta(y_i)} \ln\frac{-1 + \sqrt{-\Delta(y_i)}}{-1 -\sqrt{-\Delta(y_i)}} & \Delta(y_i) < 0\,,\nonumber
\end{array}
\right.\nonumber
\end{eqnarray}
with $\Delta(y_i)  = -1 + 4 y_i$. These results are derived from~\cite{Lavoura:1992np} and agree well with the zero Yukawa limit in~\cite{Joglekar:2012vc}.

In figure~\ref{fig:ds} we show the oblique corrections to $S$ as a function of $m_{N_4}$ and $\Delta m$: they are negligibly small in all the considered mass range and for a broad spectrum of mass splittings. On the contrary, note from figure~\ref{fig:dt} that $\Delta T$ is sensitive to the mass splitting between $E_4$ and $N_4$ only, and tends to zero for a degenerate doublet. We conclude that electroweak precision data do not constrain the mass range for $m_{E_4}$, while they severely restrict the mass splitting between the neutral and charged component, which can be at most 65 GeV at $3\sigma$.

\subsection{Fourth Generation Leptons and Collider Searches}

The nature of the vector-like doublet $L_4$ makes it loosely constrained by colliders; the drawback, however, is that it is elusive as far as it concerns its detection as well. The imposed $Z_2$ symmetry implies that in a collider the 4th generation leptons are produced always in a even number. The most probable processes are (i) pair of charged fermions ($E_4^-E_4^+$) through the exchange of $\gamma,Z$ bosons, (ii) combination of charged fermions plus its neutral partner $E_4^{\pm}N_4$ via the exchange of a $W$ boson. At LHC the $W$ production is larger than the production of $Z$ bosons and the pair creation via the process $q \bar{q} \to Z \to N_4 \bar{N}_4$ is reduced by almost two orders of magnitude with respect to the production of a charged lepton plus its companion neutrino~\cite{Carpenter:2010bs}. Therefore the dominant production rate of $L_4$ particles is through $W$ boson, namely via the process $qq' \to W \to E_4 N_4$. Because there is no mixing with the SM fermionic families, $E_4$ will decay through the process $E_4 \to N_4 W$; on the other hand we recall that the 4th generation neutrino is stable.

In case of pair production the whole process is $pp \to E_4^+ E_4^- \to N_4 \bar{N}_4 W^+ W^-$; subsequently the possible final states are
\begin{enumerate}
\item one lepton + di-jet and missing energy,
\item two oppositely charged leptons and missing energy,
\item 4 di-jet + missing energy,
\end{enumerate}
depending on whether the $W$s decay hadronically (most probable) or not.

In case the charged particles are produced along with its neutral partner the complete process at LHC is $pp \to E_4 N_4 \to N_4 N_4 W$. This results in
\begin{enumerate}
\item di-jets + missing energy,
\item single lepton + missing energy.
\end{enumerate} 
These final states do not rely on a particular signature rather it will be lost in the huge $W$ background at LHC. Usually 4th generation of leptons are supposed to produce like sign di-lepton signals, which can be well separated from the background with the opportune cuts, however this holds only if the neutrino is unstable and decays into the detector~\cite{Carpenter:2010sm,Rajaraman:2010ua}. Although $N_4$ escapes undetected at colliders, it can be probed by DM direct searches. Constraints on a 4th generation of lepton from LHC data are beyond the scope of this paper, however we remark that these might be carried out in a similar way as constraints on extra dimension have been sets by means of searches of exotic decays of $W$ bosons, see {\it e.g.}~\cite{:2012gf,CMS-PAS-EXO-11-061}.

\section{Conclusions}\label{sec:concl}
In summary, we presented a simple extension of the standard model by the addition of a vector-like massive lepton doublet, where the neutral member of the doublet $N_4$ can play the role of a dark matter, if it has a small Majorana mass. Both the asymmetry in the lepton and dark matter sector are generated simultaneously via out-of-equilibrium decay of triplet scalars via type-II leptogenesis. The model seems to satisfy all cosmological as well as laboratory constraints and has the potential to explain the current dark matter search results. Such models could also be theoretically motivated by grand unified theories such as $E_6$.

\section*{Acknowledgements}
One of the authors (RNM) would like to thank Z. Chacko for some discussions. The work of RNM has been supported by the National Science Foundation grant No. PHY-0968854. CA is supported by a European Council Starting Grant, under grant agreement No. 277591, PI G. Bertone.

\section*{References}
\bibliographystyle{elsarticle-num.bst}
\bibliography{biblio_1}

\end{document}